\documentclass{article}

\def\mytitle#1{\setcounter{equation}{0}
\setcounter{footnote}{0}
\begin{flushleft}\Large\textbf{#1}\end{flushleft}
\vspace{0.25cm}}
\def\myname#1{\leftline{{\large #1}}\vspace{-0.13cm}}
\def\myplace#1#2{\small\begin{flushleft}\textit{#1}\\
\texttt{#2}\end{flushleft}}
\newenvironment{contribution}{\normalsize\noindent}{}
\def\myclassification#1{\small\noindent
Pacs no :
       #1\vspace{0.5cm}}
\usepackage{graphicx}
\begin{document}

\mytitle{Scalar tensor theory : validity of Cosmic no hair conjecture}

\vskip0.2cm
\myname{Sudeshna Mukerji\footnote{mukerjisudeshna@gmail.com}}
\vskip0.2cm
\myname{Nairwita Mazumder\footnote{nairwita15@gmail.com}}
\vskip0.2cm
\myname{Ritabrata Biswas\footnote{biswas.ritabrata@gmail.com}}
\vskip0.2cm
\myname{Subenoy Chakraborty\footnote{schakraborty@math.jdvu.ac.in}}
\vskip0.2cm
\vskip0.2cm
\myplace{Department of Mathematics, Jadavpur University, Kolkata-32, India.}{}
\vskip0.2cm

\myclassification{}

\begin{abstract}
The paper deals with cosmic no hair conjecture in scalar tensor theory of gravity. Here we have considered both Jordan frame and Einstein frame to examine the conjecture. In Jordan frame, one should restrict both the coupling function of the scalar field and the coupling parameter in addition to the ususal energy conditions for the the matter field for the validity of CNHC while in Einstein frame the restrictions are purely on the energy conditions.
\end{abstract}

\begin{contribution}
\end{contribution}
\section{Introduction}
In general terminology, the Cosmic No Hair Conjecture(CNHC)
states that "all expanding universe models with a positive
cosmological constant asymptotically approach the de Sitter
solution." To address the question of whether the universe
evolves to a homogeneous and isotropic state during an
inflationary epoch, Gibbons et al (1977) \cite{Gibbons1} and Hawking et al (1982)
\cite{Hawking1} developed this conjecture. Subsequently Wald (1983) \cite{Wald1} gave a formal
proof of it for homogeneous cosmological models (Bianchi models)
with a positive cosmological constant. He assumed that the matter
field should satisfy the strong and weak energy conditions.
Kitada and Maeda (1992) \cite{Kitada1} proved a cosmic no hair theorem for
Bianchi models in power-law inflation. They also showed that the
conclusion is true for Bianchi type IX if the initial ratio of the
vacuum energy to the maximum three curvature is larger than
one-half. Then, Chakraborty and Paul (2001) \cite{Chakraborty1} studied the CNHC for
anisotropic Bianchi models which admit an inflationary solution
with a scalar field. They found that the form of the potential
does not affect the evolution in the inflationary era while the
late time behaviour is controlled by the constant additive factor
in the potential for the inflaton field.  Chakraborty and
Chakraborty (2002) \cite{Chakraborty2} have studied the CNHC  for homogeneous
anisotropic Bianchi models with a varying cosmological
constant \cite{Vishwakarma1, Vishwakarma2} in Randall-Sundrum braneworld type scenarios.
They showed that in the first case the universe will isotropize
after power-law inflation while there is exponential expansion in
the second case. Chakraborty and Debnath (2003) \cite{Chakraborty3} have shown using
examples with realistic fluid models that strong and weak energy
conditions are sufficient for the CNHC  in braneworld scenarios.
Later, Chakraborty and Bandopadhyay (2007)\cite{Chakraborty4} have showed that for
validity of CNHC, in brane scenarios, the matter in the bulk need
not be standard matter but must obey some restrictions depending
on the brane tension. Then they (2008)\cite{Chakraborty5} have proved the CNHC for
Gauss-Bonnet dilatonic scalar coupled to Einstein gravity with
coupling parameter growing linearly in time. In this paper, we
attempt to extend Walds' \cite{Wald1} result in the context of scalar tensor theory and
compare the restrictions for the validity of CNHC with those
in general relativity.

Soon  after the introduction of general relativity (GR) several attempts were there to form alternative theories of gravity. Some interesting alternatives were based on an effort to build a more general theory by dropping one or more of the several assumptions of GR, like the fact that the only degrees of freedom of the gravitational field are those of the metric, or the simplicity of choice that the gravitational Lagrangian should be a linear function of the scalar curvature. One of the most studied alternative theories is scalar tensor theory, where the gravitational action contains, appart from the metric a scalar field which describes the part of the gravitational field.

ususally, there are two types of frames : the Jordan frame and the Einstein frame. The scalar tensor theory can be formulated in the Jordan frame where the scalar field $\phi$ is coupled nonminimally to the Ricci scalar $R$ but not directly to the matter, whereas the scalar field kinetic term involves an arbitrary function $F(\phi)$. In Einstein frame the scalar field is minimally coupled to the Ricci scalar and its kinetic term is in the canonical form. In this paper we consider scalar tensor theory both in the Jordan frame and in the Einstein frame in the sections 2 and 3 respectively and try to examine the validity of the cosmic no hair conjecture for both the frames.

\section{Cosmic no hair conjecture in Jordan frame}
In the four dimensional space time the action for the scalar tensor theory in the Jordan frame is given by
$$
S=\int
d^{4}x\sqrt{-g}\left[\frac{1}{2\kappa^{2}}F\left(\Phi\right)R-\frac{3\left(1-4\beta^{2}\right)}{16\kappa^{2}\beta^{2}}\frac{1}{F\left(\Phi\right)}\left(\frac{dF\left(\Phi\right)}{d\Phi}\right)^{2}\left(\nabla\Phi\right)^{2}-V\left(\Phi\right)\right]+\int
d^{4}x\sqrt{-g}L_{m}\left(g_{\mu\nu}\right)
$$
where $\kappa^{2}=8\pi G$, $\beta$ is a coupling constant, $g$ and
$R$ are the determinant and the curvature scalar of the metric
tensor $g_{\mu\nu}$ respectively, $L_{m}$ is the matter
Lagrangian, $F>0$ is a function of the scalar field $\Phi$ and
$V(\Phi)$ is the potential for the scalar field.

Here matter is chosen in the form of perfect fluid having energy momentum tensor $$T_{\mu\nu}^{(m)}=\left(\rho_{m}+p_{m}\right)u_{\mu}u_{\nu}+p_{m}g_{\mu\nu}$$
where $u_{\nu}$ is the fluid four velocity.

Now varying the action over the field variables the field equations can be written as the effective Einstein equations
\begin{equation}\label{1}
G_{\mu\nu}=\kappa^{2}\left[T_{\mu\nu}^{eff}+T_{\mu\nu}^{(m)}\right],
\end{equation}
and the equation for the scalar field (i.e., the wave equation for the scalar field)
\begin{equation}\label{2}
\frac{d^{2}F}{dt^{2}}+3H\frac{dF}{dt}=\frac{4\beta^{2}}{3}\left[4V(\Phi)-2\frac{dV(\Phi)}{d\Phi}\frac{F}{\frac{dF}{d\phi}}+\left(\rho_{m}-3p_{m}\right)\right]
\end{equation}
where $T_{\mu\nu}^{eff}$ is the effective enrgy-momentum tensor which arises due to additional terms for the coupled scalar field.

The continuity equation namely
\begin{equation}\label{3}
\frac{d\rho_{m}}{dt}+3H\left(\rho_{m}+p_{m}\right)=0,
\end{equation}
is same as in Einstein gravity.

Now, in order to examine the cosmological evolution, we start with the initial value constraint (i.e., Hamiltonian constraint)
\begin{equation}\label{4}
G_{\mu\nu}n^{\mu}n^{\nu}-\kappa^{2}\left[T_{d(n)}^{eff}+T_{d(n)}^{(m)}\right]=0,
\end{equation}
and the RayChaudhuri equation
\begin{equation}\label{5}
R_{ab}n^{a}n^{b}-\kappa^{2}\left[T_{s(n)}^{eff}+T_{s(n)}^{(m)}\right]=0,
\end{equation}
where, $n^{a}$ is the unit normal to the homogeneous hypersurface and $H$ is the Hubble parameter. The expression for the matter components are
$$T_{d(n)}^{(l)}=T_{\mu\nu}^{(l)}n^{\mu}n^{\nu}~~and~~ T_{s(n)}^{(l)}=\left(T_{\mu\nu}^{(l)}-\frac{1}{2}Tg_{\mu\nu}\right)n^{\mu}n^{\nu}$$
with $l=(eff,~m)$. One may note that positivity of $T_{d(n)}^{(l)}$ and $T_{s(n)}^{(l)}$  implies the validity of dominant and strong energy condition.

In the standard $(3+1)$ decomposition of the 4-dimensional manifold, the metric of the hypersurface (3-space) and the extrinsic curvature have the expressions
\begin{equation}\label{6}
q_{ab}=g_{ab}+n_{a}n_{b}
\end{equation}
and
\begin{equation}\label{7}
K_{ab}=\frac{1}{3}Kq_{ab}+\sigma_{ab}
\end{equation}
where $K=K_{ab}q^{ab}$ is the trace of the extrinsic curvature and $\sigma_{ab}$ is the shear of the time like geodesic congruence orthogonal to the homogeneous hypersurfaces. The dynamical equations (\ref{4}) and (\ref{5}) can be expressed in terms of the three space variables as
\begin{equation}\label{8}
K^{2}=3\kappa^{2}\left[T_{d(n)}^{eff}+T_{d(n)}^{(m)}\right]+\frac{3}{2}\sigma_{ab}\sigma^{ab}-\frac{3}{2}^{(3)}R
\end{equation}
and
\begin{equation}\label{9}
\dot{K}=-\kappa^{2}\left[T_{s(n)}^{eff}+T_{s(n)}^{(m)}\right]-\sigma_{ab}\sigma^{ab}-\frac{1}{3}K^{2}
\end{equation}
where $^{(3)}R$ is the scalar curvature of the homogeneous
hypersurface and is shown to be negative for Bianchi type
homogeneous anisotropic cosmological models. Here $'.'$ denotes
the Lie derivative with respect to proper time.

Thus approaching along the idea of Wald \cite{Wald1}, we find that for validity of CNHC the combination of the ordinary matter and the effective matter should satisfy dominant and strong energy conditions,i.e.,
$$T_{d(n)}^{eff}+T_{d(n)}^{(m)}\geq 0 ~~and~~T_{s(n)}^{eff}+T_{s(n)}^{(m)}\geq 0.$$
In explicit form the above inequalities can be written as
\begin{equation}\label{10}
\rho_{m}+V(\Phi)-H\frac{dF}{dt}+\frac{1-4\beta^{2}}{16\beta^{2}}\frac{1}{F}\left(\frac{dF}{dt}\right)^{2}
\geq 0
\end{equation}
and
\begin{equation}\label{11}
\left\{\rho_{m}\left(1+4\beta^{2}\right)+3p_{m}\left(1-4\beta^{2}\right)\right\}-\frac{\left(1-4\beta^{2}\right)}{\beta^{2}}\frac{1}{F}\left(\frac{dF}{dt}\right)^{2}-\frac{2}{3}\left(1-8\beta^{2}\right)V(\Phi)-10H\frac{dF}{dt}-8\beta^{2}\frac{FV'}{F'}\geq
0
\end{equation}
The above inequalitites are satisfied provided

(a)$F$ is decreasing function of $\Phi$,i.e., $F'=\frac{dF}{d
\Phi}<0$

(b) The coupling parameter $\beta$ is restricted to the range $\frac{1}{8}<\beta^{2}<\frac{1}{4}$

(c) The ordinary matter should satisfy both the dominant and the strong energy conditions,i.e., $\rho_{m}\geq 0$ and $\rho_{m}+3p_{m}\geq 0$.

\section{Scalar Tensor Theory in Einstein frame and CNHC}
The action in the Einstein frame takes the form
$$
S=\int d^{4}x\sqrt{-\tilde{g}}\left[\frac{1}{2\kappa^{2}}\tilde{R}-\frac{1}{2}\left(\tilde{\nabla}\phi\right)^{2}-\tilde{V}(\phi)\right]+\int d^{4}x \sqrt{-\tilde{g}}\tilde{L}_{m}\tilde{g}_{\mu\nu}exp\left\{-2\sqrt{\frac{2}{3\kappa}}\beta\phi\right\}
$$
where
$g_{\mu\nu}=\tilde{g}_{\mu\nu}exp\left\{-2\sqrt{\frac{2}{3\kappa}}\beta\phi\right\}$
, $F(\Phi)=exp\left\{2\sqrt{\frac{2}{3\kappa}}\beta\phi\right\}$
and $V(\phi)=F(\Phi)^{2}\tilde{V}\left(\phi\right)$.

Now proceeding as in the previous section, the dynamical equations can be expressed as
\begin{equation}\label{12}
K^{2}=\frac{3}{2}\sigma_{\mu\nu}\sigma^{\mu\nu}-\frac{3}{2}^{(3)}R+\kappa^{2}\left[\tilde{\rho}_{m}+\tilde{\rho}_{\phi}\right]
\end{equation}
and
\begin{equation}\label{13}
\dot{K}=-\frac{1}{3}K^{2}-\sigma_{\mu\nu}\sigma^{\mu\nu}-\frac{\kappa^{2}}{6}\left(\tilde{\rho}_{m}+3\tilde{p}_{m}\right)-\frac{\kappa^{2}}{6}\left[\left(\tilde{\rho}_{m}+3\tilde{p}_{m}\right)+\left(\tilde{\rho}_{\phi}+3\tilde{p}_{\phi}\right)\right]
\end{equation}
and the matter conservation equations becomes
\begin{equation}\label{14}
\frac{d\tilde{\rho}_{m}}{d\tilde{t}}+3\tilde{H}\left(\tilde{\rho}_{m}+\tilde{p}_{m}\right)=-\sqrt{\frac{2}{3}}\kappa\beta\frac{d\phi}{d\tilde{t}}\left(\tilde{\rho}_{m}-3\tilde{p}_{m}\right)
\end{equation}
where
\begin{eqnarray}\label{15}
\nonumber d\tilde{t}=exp\left\{\sqrt{\frac{2}{3}}\kappa\beta\phi\right\}dt~~, ~~\tilde{p}_{m}=exp\left\{-4\sqrt{\frac{2}{3}}\kappa\beta\phi\right\}p_{m}~~,\\ \tilde{\rho}_{m}=exp\left\{-4\sqrt{\frac{2}{3}}\kappa\beta\phi\right\}\rho_{m}~~,~~\tilde{H}=exp\left\{-4\sqrt{\frac{2}{3}}\kappa\beta\phi\right\}H.
\end{eqnarray}
Here $\tilde{\rho}_{\phi}=\left\{\frac{1}{2}\left(\frac{d\phi}{d\tilde{t}}\right)^{2}+\tilde{V}(\phi)\right\}$ , $\tilde{p}_{\phi}=\left\{\frac{1}{2}\left(\frac{d\phi}{d\tilde{t}}\right)^{2}-\tilde{V}(\phi)\right\}$ are the effective energy density and effective pressure to the scalar field.

The transformation (\ref{15}) shows that the equation of state for the matter remains same both in Einstein frame and in Jordan frame, hence energy conditions are not affected by the frame transformation. Thus, validity of CNHC demands that both the energy conditions (namely dominant and strong) are satisfied by the matter field and the effective scalar field in Einstein frame. One may note that the above results in Einstein frame are similar to those in brane world scenario.

Therefore, compared to Einstein gravity, we need some restrictions on the scalar field for the validity of the CNHC in scalar-tensor theory.

\begin{contribution}

{\bf Acknowledgement :}

SM is thankful to Jadavpur University for allowing her to use the library and laboratory facilities. NM wants to thank CSIR, India for awarding JRF. RB is thankful to West Bengal State Govt for awarding JRF.

\end{contribution}

\end{document}